\documentclass[12pt]{article}
\usepackage{graphicx}

\newcommand{\beq}{\begin{equation}}
\newcommand{\eeq}{\end{equation}}
\newcommand{\bea}{\begin{eqnarray}}
\newcommand{\eea}{\end{eqnarray}}

\begin{document}
\begin{center}
\noindent {\Large On Multifractal Structure in Non-Representational Art}
\vskip .3cm
\noindent J.\ R.\ Mureika \\
{\footnotesize Email: jmureika@lmu.edu \\
{\it Department of Physics, Loyola Marymount University, Los Angeles, CA~~90045-8227}}\\
\noindent C.\ C.\ Dyer \\
{\footnotesize Email: dyer@astro.utoronto.ca \\
{\it Department of Astronomy and Astrophysics, University of Toronto, Ontario, Canada}}\\
\noindent G.\ C.\ Cupchik \\
{\footnotesize Email: cupchik@scar.utoronto.ca \\
{\it Division of Life Sciences, University of Toronto at Scarborough, Ontario, Canada}}
\end{center}
\vskip .5cm
\noindent Abstract\\ 
Multifractal analysis techniques are applied to patterns in several 
abstract expressionist artworks, paintined by various artists.
The analysis is carried out on two distinct types of
structures: the physical patterns formed by a specific color (``blobs''),
as well as patterns formed by the luminance gradient between adjacent colors
(``edges'').  It is found that the analysis method applied to ``blobs'' 
cannot distinguish between artists of the same movement, yielding a 
multifractal spectrum of dimensions between about $1.5-1.8$. The method
can distinguish between different types of images, however, as demonstrated
by studying a radically different type of art.   The data suggests that
the ``edge'' method can distinguish between artists in the same movement, 
and is proposed to represent a toy model of visual discrimination.  A
``fractal reconstruction'' analysis technique is also applied to the images,
in order to determine whether or not a specific signature can be extracted
which might serve as a type of fingerprint for the movement.  However,
these results are vague and no direct conclusions may be drawn.

\vskip 1cm
\noindent PACS Primary: 89.75.Da; secondary: 95.75.mN; tertiary: 89.75.Kd
\pagebreak

\section{Introduction and Background}
The use of fractal analysis methods to study structure in art and music
is not a new field.  Recently, the question of perceptability of such fractal
structure has been addressed.  The authors of 
References~\cite{schmuckler1,schmuckler2} pose the question of whether or not
humans are ``attuned'' to the perception of fractal-like
optical and auditory stimuli.  Similarly, \cite{gliden}
suggests that there is a fractal-like signature in memory processes which
can be detected in the statistical variance of averaged repeated actions
(such as repeated drawing lines of specific lengths or shapes; the statistical
variations in the lengths are shown to be not purely random noise, but 
fractally ordered ``$1/f$'' noise).  

In the visual arts, there have been several contributions made by the
authors of \cite{taylor,taylor1,taylor3,taylor2}, 
in which the paintings of Jackson Pollock play a prominent role. Of their
many interesting conclusions, the most striking is that Pollock's drip
paintings almost uniformly possess a fractal dimension around 1.7.
This was confirmed by the authors of \cite{jrmths,jrmleo}, who also extended
the study to other artists of the abstract expressionist school
(notably the Qu\'{e}bec-based group Les Automatistes).  It
was discovered that many of the paintings from these other artists
possess a similar fractal dimension.

Similar research on computer-generated cosmological models suggests that 
the fractal (or box) dimension 
is a vague statistic for identifying structural differences in point
sets, and that the full multifractal spectrum can yield deeper information
as to the nature of the distributions \cite{jrmccd}.  
However, it has been concluded that
the utility of the method seems limited to identifying {\it classes} of
distributions formed by different mechanisms, and not differences between 
individual members of the same class.  

In the following paper, this
method will be applied to two-dimensional, non-representational images, to
ascertain whether or not similar statements may be made about the analysis.
Summary results of this study have been reported in Reference~\cite{jrmleo}, but
this paper will greatly expand upon the data and present technical details of
the analysis in diverse ways.  An overview of fractal and multifractal theory
is first presented.  Following this, the fractal (box) dimension for several
abstract expressionist paintings by different artists is performed, and this
is contrasted with the information dimension for the same works.
The box method is tested for robustness in Section~\ref{robust}. 

As a control, these abstract expressionist are contrasted
with the deterministic ``Artonomy'' paintings by Tsion Avital \cite{avital}.
Like the three dimensional distributions, the two former works can be 
interpreted to be formed by one type of mechanism (although the comparison
is perhaps more ambiguous), while the latter
a decidedly different mechanism (this differentiation will be further 
discussed in Section~\ref{artonomy}).  

The analysis will then be extended to include the full multifractal spectrum
for each work.  A comprehensive analysis is performed on color patterns
for the artists in question, beginning in Section~\ref{multiart}.  
Furthermore, issues addressing potential image reconstruction
from the multifractal spectrum is discussed in Section~\ref{uses}.
Finally, as a toy model for visual discrimination
based on the notion that human perception may be influenced by contrast
edges instead of colors, the multifractal spectra of contrast patterns
in the paintings are analyzed in Section~\ref{visual}.   Some limitations
of the method are discussed in Section~\ref{limitations}, and concluding 
remarks are summarized in Section~\ref{conclusions}.

\section{Theory of Multifractals}
\label{multi}
The similarity in form and function
of the classic fractal (or box) dimension, Shannon's information 
dimension, and the statistical correlation dimension is
not a coincidence (see References~\cite{mandel1,falconer,barnsley,inftheory1}
for general details on these dimensions).  
In fact, these quantities are but three members
of an {\it infinite} set of dimensions which characterize a fractal set.
Since first being introduced as a method of describing or quantifying
the behavior of
strange attractor sets and turbulent flows,
multifractal analysis has gained steady momentum in physics and fields abroad
\cite{proc1,proc2,grass,halsey1,proc4}.
Comprehensive reviews of multifractals and their applications in the physical
sciences are available in the aforementioned references, as well
as such works as \cite{tel,paladin,vicsek}, in addition to
many of the references cited hereafter.
                                                                                
Classic geometric ``monofractals'' such as the Koch snowflake or Sierpinski carpet 
are defined by a single scale invariant behavior,
which of course is the fractal dimension, but in many cases a single such
power law fails to characterize completely the distribution in question.
Multifractals, on the other hand, may be regarded as
an intricate weave of an infinite number of fractals, all of which are
characterized by different (local) scaling dimensions.  That is, each
subset forms a ``sub-fractal'' describing a distinct sub-structure of
the whole.  It is much
more reasonable and realistic to expect natural objects to exhibit this
behavior.
                                                                                
Multifractal dimensions are generalizations of the Hausdorff measure
\cite{falconer}.  The partition function for an
$\epsilon$-covering ({\it i.e.} balls of radius $\epsilon_i$)
is defined as \cite{proc2,proc4}
\beq
\Gamma(q,\tau) = \sum_i \frac{p_i^q}{\epsilon_i^\tau}~,
\label{genhaus}
\eeq
with $p_i$ a measure of the set or pattern density in the ball.
For given $q,\tau(q) \in {\cal R}$, take the supremum (or infimum, depending
on whether $q$ is positive or negative, respectively) in the limit
$\epsilon_i \rightarrow 0$ (thus find the minimal covering set for
the generalized measure).  Then,
there exists a critical value of $\tau \equiv \tau(q)$ for which
$\Gamma(q,\tau)$ goes from convergence to 0, to divergence to $\infty$.
At this transitionary value, the sum converges
as $\Gamma(q,\tau) = {\rm const}$.  The minimal covering may be generalized
to balls of equal radii $\epsilon_i = \epsilon$, whence it follows that
\beq
\Gamma(q,\tau) = \epsilon^{-\tau(q)} \sum_i p_i^q \sim 1~,
\eeq
for a suitable renormalization of the measure.  Hence,
\beq
\sum_i p_i^q \sim \epsilon^{\tau(q)}~,
\eeq
and thus in the limit $\epsilon \rightarrow 0$, it can be shown that
\beq
\tau(q) = \lim_{\epsilon \rightarrow 0}
\frac{\log[\sum_i p_i^q]}{\log\epsilon}~.
\label{mfdef}
\eeq
Let the measure partition function over $N(\epsilon)$ balls of radius $i$ be
\beq
Z(q,\epsilon) = \sum_{i=1}^{N(\epsilon)} \; [p_i(\epsilon)]^q~,
\label{zpart}
\eeq
and define the general scaling relation
$Z(q,\epsilon) \sim \epsilon^{(q-1)\;D_q}$
\cite{vicsek}, which ensures recovery of the Hausdorff dimension
for $q=0$, as well as normalization of (\ref{zpart}) for $q=1$.
From (\ref{mfdef}), it can be concluded that
\beq
D_q = \frac{\tau(q)}{q-1}~.
\label{dqdef}
\eeq
In the limit $q \rightarrow 0$, (\ref{genhaus}) reduces to the usual box
dimension.  Furthermore, the information and
correlation dimensions are recovered in the limits $q \rightarrow 1,2$.

In practical applications, the box counting method can be generalized
to obtain the values of $\tau(q)$ for given $q$.
That is, modify the partition function $Z(q,\epsilon)$ over $N(\epsilon)$
to sets of covering boxes (instead of balls) of equal side $\epsilon$,
where as before $p_i(\epsilon)$ is the relative density of the
set in box $i$.
The scaling information of each moment is obtained by taking the
logarithmic derivative of (\ref{zpart}) with respect to box size, where
\beq
\tau(q) = \frac{d\log[Z(q,\epsilon)]}{d\log(\epsilon)}~.
\label{taudef}
\eeq
                                                                                
The moment parameter $q$ can be thought of as a filter, which identifies only
the singularity characteristics of the distribution at a particular
``degree'' of clustering.  Increasing values of $q > 0$ emphasize the
stronger local clustering nature of the pattern, while decreasing values
of $q < 0$ the less singular regions.   That is,
higher values of $q$ serve to ``eliminate'' the smaller values of $p_i$,
yielding a subset of the overall distribution whose scaling behavior is
more condensed (and vice versa for negative $q$).
Likewise, the other $D_q$ provide a measure of the number of q-tuples
whose mutual separation is contained within a covering box (ball) of 
size $\epsilon$.  Thus, a quantitative measure of the $D_q$
spectrum yields an understanding of all order of correlations amongst
clusters of varying densities.
                                                                                
Certain key values of $D_q$ are extremely useful in characterizing the physical
clustering characteristics of a set.  In addition to the values
$D_{0,1,2}$ mentioned before, the generalized dimensions for the
limits $q\rightarrow \pm \infty$ yield valuable information about the
maximal and minimal density regions of the set.
$D_{\infty}$ is a measure of the scaling behavior for
the densest clustering regions of the multifractal,
while $D_{-\infty}$ corresponds to
the equivalent for the least dense or ``rarefied'' regions.
                                                                                
In essence, the multifractal measures give an indication of ``how fractal
is the fractal''.   A measure of the difference between the box dimension
and any successive $D_q$ value
\beq
\Delta = D_q - D_0 ~.
\label{diffdq}
\eeq
provides an estimate of the ``degree'' of inhomogeneity of the associated
probability distribution \cite{paladin}.  In particular, it seems reasonable
to evaluate $\Delta|_{q \rightarrow \infty}$ as an overall gauge of the
``depth'' of inhomogeneity.  Clearly, it follows that
$\Delta = 0$ for single-scaling Euclidean normal or monofractal sets, so
the larger the value of $\Delta$, the greater the ``multifractality''.
                                                                                
\section{Fractal Expressionism}
\label{previous}
In the last 1990s, the application of fractal analysis to the study of 
abstract expressionist art began to gain momentum, the first of which
was reported in \cite{taylor1}.  It was 
concluded that Jackson Pollock's work did indeed present certain fractal 
characteristics.  Coined ``Fractal Expressionism'', the authors in question 
proposed
that Pollock's drip paintings were constructed by processes not unlike those
which help to forge the myriad of similarly fractal natural phenomena.
In fact, they further suggested that by painting with such ``automatism'',
Pollock succeeded in capturing the very essence of nature within his works.

In particular, it was noted that most of the paintings studied contain at
least two distinct scaling behaviors at different levels, much the same
as the debated transitions to homogeneity in galaxy clustering.  The first
of these occurred at scales on the order of 1~mm to about 5~cm, beyond
which point a second scaling is observed up to scales of several meters
\cite{taylor}.  After a review of Pollock's painting methods and techniques,
it was determined that these two dimensions were the result of two distinct
physical processes.  The larger-scale patterns resulted from Pollock's
``Levy flights'' across the canvas (a Levy Flight is a combination of
discrete, random jumps coupled with local fractal Brownian motion
\cite{vicsek}).  Likewise, the small-scale structure was attributed to his
infamous ``drip'' technique, which was largely dependent on the physical
characteristics of the paint (viscosity, the height from which it was
dripped, absorption into the canvas, {\it etc...}).  

These two dimensions are coined $D_L$ (Levy) and $D_D$ (Drip), and in
general it was found that $D_L > D_D$ (in fact, the authors of 
\cite{taylor} claim that $D_L$ tended to values close to 2, indicative of
the ``space-filling'' behavior of Pollock's Levy Flights).  Furthermore,
it was claimed that $D_L$ tended to increase from 1 to about 1.7 between
the early 1940s to the late 40s / early 50s (around the time Pollock perfected
his drip technique \cite{pollock1}).

Their analysis focused exclusively on the works of Jackson Pollock, and
these dimensions are attribute to his own artistic style.
In the spirit of the aforementioned conclusions of \cite{jrmccd}, however, 
the question should
be asked as to whether or not such an analysis truly pinpoints anything
``unique'' about the artist in question, or whether the resulting 
statistics are shared by a common set of images and patterns formed by 
similar methods.  

\subsection{Image specifics}
\label{iamgespecs}
The majority of the images considered in this study are digital scans of
Pollock's works from the references \cite{pollock1, pollock2}.  Images
by Les Automatistes have been scanned from \cite{autobook1}.  The resolution
of the scans was chosen as 300~dpi, creating images roughly 1000~pixels (px) in
length (longest side) and files 20~Mb in size.  The analysis has been performed
on approximately 25 Jackson Pollock paintings, revealing similar trends for
each.  However, this discussion will be restricted to a small sample set of 
six.  The images herein are listed in Table~\ref{appendix1}.
At the specified resolution, each pixel corresponds to approximately
0.1-0.4~cm, although this will depend on the actual reduction scale from 
the base image. 

The covering boxes range in size from $d = $1024~px to $d =$4~px, 
or length scales of roughly $1.5-2.5~$m to a few millimeters.  
Hence, the analysis covers about 3 orders of magnitude.  Higher 
resolutions could allow for greater range of scales, but would correspond 
to much larger images and longer run-times / higher memory requirements
for the code.  It was verified
that the quality of the fits did not change appreciably for a lower limit
of $d=2$, and the estimated dimensions were statistically equal to within the
associated error.

\subsection{Color Variance Filter Process}
\label{cutoff}
Accurate definitions of colors and color differences are very difficult to
obtain.  Any investigation which relies on color matching must do so with
care.  The following procedure is a rough example of how like colors might be
extracted from an image, based on their Euclidean separation in the three
dimensional RGB color space.

To trace or filter the pattern
of a given pigment, the variation in shading is accounted for via the
color-variance filter process.  The images studied herein are 24-bit color
maps, hence each separate channel may assume 256 possible values.  
An RGB triplet is chosen as the target
color, each pixel (channel) intensity in the image
is then compared to the initial triplets R$_0$, G$_0$, B$_0$ (hereafter
RGB$_0$), and the Euclidean distance (or color radius) is calculated,
\beq
R_{RGB} = \sqrt{(R_0 - R_{pix})^2 + (G_0 - G_{pix})^2 + (B_0 - B_{pix})^2}~,
\label{colorradius}
\eeq

\begin{figure} \begin{center} \leavevmode
\includegraphics[width=0.5\textwidth]{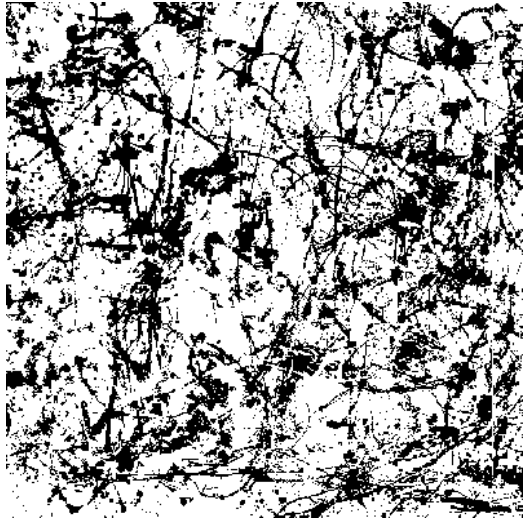}
\end{center} \caption{
Black pigment filtered image P02, $\beta \sim 20$.
}
\label{ARblob}
\end{figure}

\begin{figure} \begin{center} \leavevmode
\includegraphics[width=0.5\textwidth]{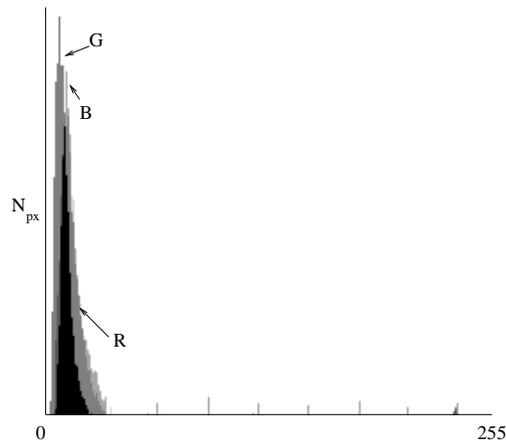}
\end{center} \caption{
Pixel count $N_{\rm px}$ histogram for black pattern of P04,
showing peaks at $(\rm{R}, \rm{G}, \rm{B}) = (21, 17, 21)$.
}
\label{p32blackhist}
\end{figure}

Figure~\ref{ARblob} shows the filtered pattern for $\beta_{RGB} = 20$
for image P02.
Patterns are isolated by including pixels for which $R_{RGB} \leq \beta$, a
cutoff whose value is determined by examination of the RGB histogram
for the color in question.   
Figure~\ref{p32blackhist} shows the R, G, and B pixel intensity histogram
for the ``black'' pigment of image P04, which is generally of the same
form for all images considered herein.
The peaks of each correspond roughly
to the values $(\rm{R}, \rm{G}, \rm{B}) = (21, 17, 21)$, which is 
taken as the target color RGB$_0$.  Note the smooth drop-off
for increasing (and decreasing) values of the pixel intensities.  

For the paintings considered, it was found that most RGB histogram spreads
tend to extend no more than 5-20 pixel intensities from the central peak.
Hence, it seems reasonable to assume that the cutoff $\beta$ should be between
$\beta \in (10\sqrt{3}-20\sqrt{3}) \simeq (10, 40)$.  

The pseudo-normal nature of the distribution in Figure~\ref{p32blackhist}
suggests that a Gaussian
filter, which weights colors according to their distance from the ``target'',
would be more appropriate that the cut-off filter considered presently.
This type of filter is inappropriate for calculation of the box Dimension, 
for which any box is counted in which there exists a point in the allowable range ({\it i.e.} this would result in a severe over-count of boxes).
However, a ``weighted'' information dimension is certainly feasible, in which
one assigns the color match a value of $\rm{exp}(-R^2_{\,\rm{RGB}}/a^2)$,
with $a^2$ the FWHM corresponding to the average histogram spread.
This filter would be better exploited in the multifractal analysis
of \ref{multiart}.  However, preliminary calculations suggest that the 
results will not vary significantly from those of the cut-off method described
herein.  Since the very notion of color distance discrimination itself is
somewhat of a fuzzy area (see {\it e.g.} \cite{hcv} or similar references),
it is best not to ``over-complicate'' the procedure at this given stage
of development.  Thus, only the cut-off will be used in this study.

\section{The Box and Information Dimension of Jackson Pollock's Work}
\label{polinf}
As previously mentioned, the information dimension can be
considered a better
statistic for the study of recursive patterns.  That is, the
box dimension can sometimes provide an overestimate of the scaling behavior, 
since it does not account for the relative density of points within the box.  
Although these specific results have been reported in \cite{jrmleo}, a 
slightly different analysis of the findings was given in that Reference.
What follows will be a more technical discussion of the results.

Table~\ref{poldims} presents the corresponding 
dimension estimates for each painting.  The Box Dimensions calculated by a least-squares
regression on the data points seems to provide a good agreement to the results
cited in \cite{taylor}, who found  {\it e.g.}
$D_0 = 1.67$ for P02 and 1.72 for P01.
This suggests that a value of $\beta \in [20,30]$ would be in rough agreement 
with their analysis.  

However, closer inspection of the results of Table~\ref{poldims} reveals
that in certain cases 
the estimate of the dimension is critically dependent on a correct
choice of $\beta$: the darker colors appear more stable, while the
lighter ones show wider variation.  In order for this analysis technique
to be useful, these selection criteria must be extremely well defined.
Otherwise, the results risk becoming meaningless.  Ideally, some kind
of variance in choice of 
$\beta$ should be incorporated into the overall error estimate.  The
issue of color space selection is discussed in Reference~\cite{jrmcie}.

As mentioned in the previous section, the
authors noted an apparent break in the slope of the log-log plot, and
assumed that this represented different scaling behavior of two different
mechanisms.  The shallower slope was taken to be representative of Pollock's
painting technique.  
In fact, in reference~\cite{kaye}, the author discusses the association of
two distinct dimensions based on the topological morphology of the fractal
(for higher length scales), as well as its texture (lower scales).  
These two dimensions are appropriately labeled as those of the
{\it structural fractal} and {\it textural fractal}, respectively.
Relating to the work of \cite{taylor}, it is not unreasonable to 
interpret their two dimensions accordingly, {\it i.e.} the overall
``structure'' of the painting at higher length scales,
and the fine-grained refinement at lower scales.

\begin{figure} \begin{center} \leavevmode
\includegraphics[scale=0.3,angle=270]{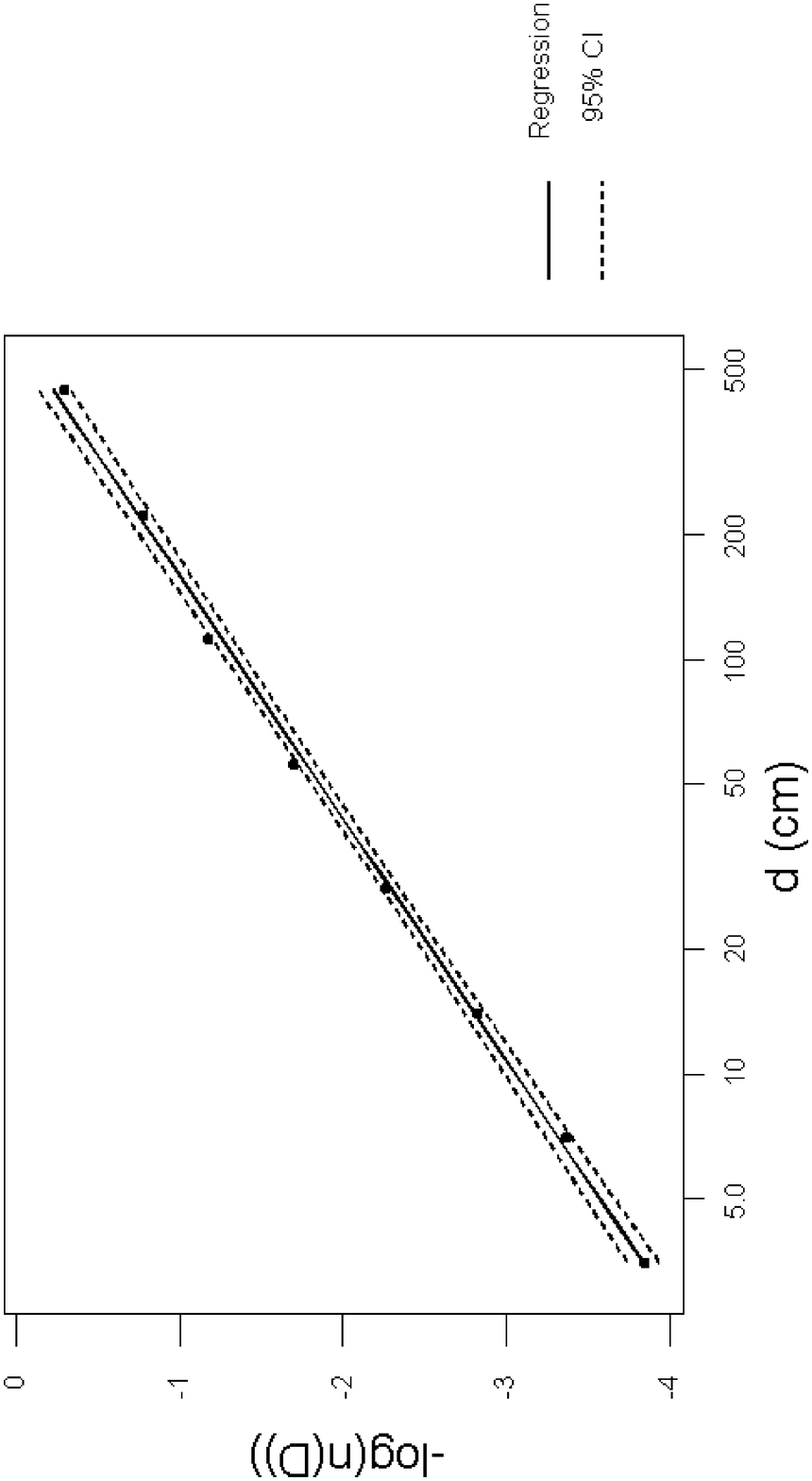}
\includegraphics[scale=0.3,angle=270]{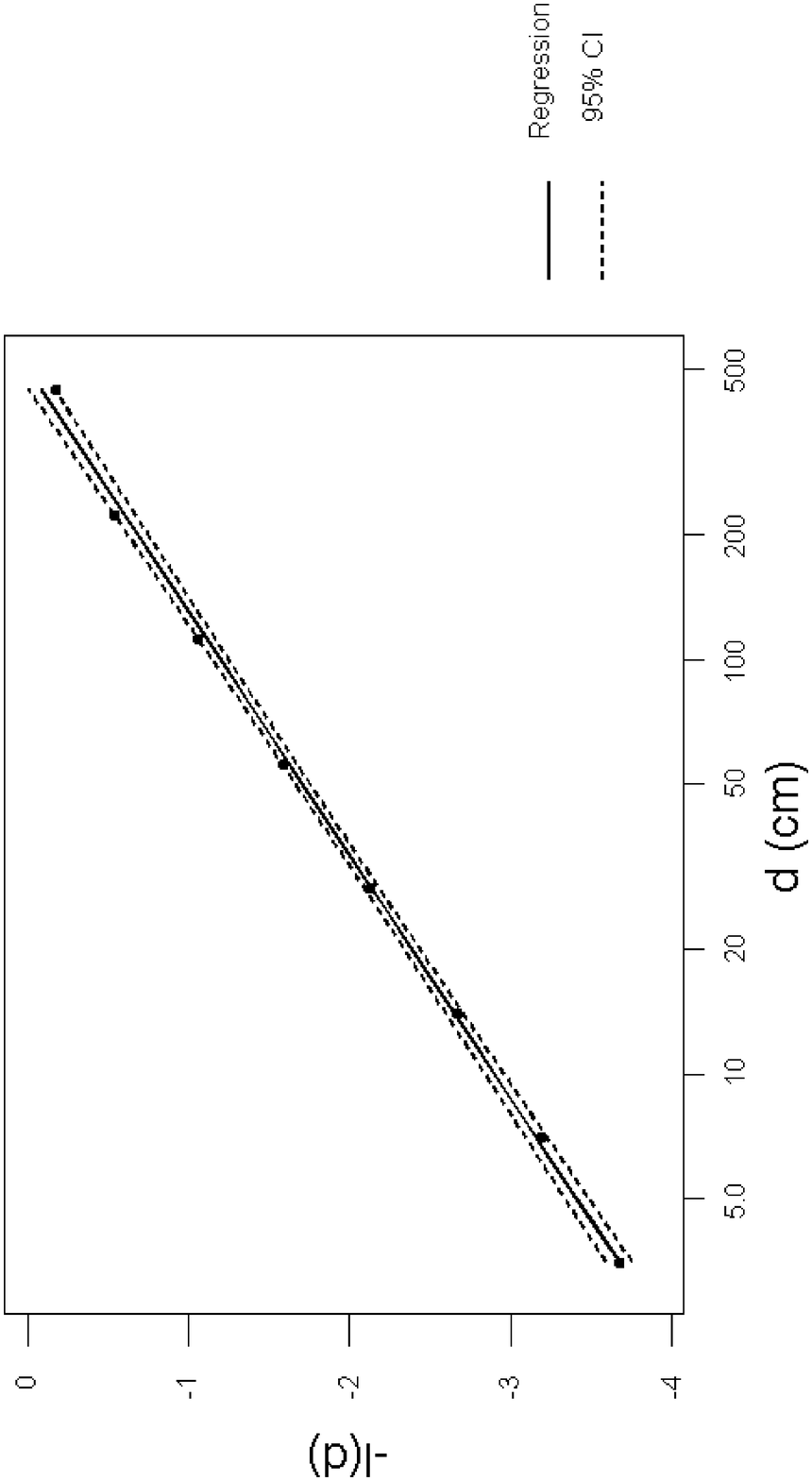}
\end{center} \caption{
Sample log-log fits used to determine box and information dimensions for P02, $\
beta = 20$.}
\label{ARcomp}
\end{figure}

Note that in the analysis presented herein,
the two-slope hypothesis of \cite{taylor} is not supported
when one considers the magnitudes of the associated errors in the
least-squares procedure.  The confidence level curves
suggest that the ``shallow'' slope at lower length scales could be explained
as statistical variation in the fit.  
The fits in Figure~\ref{ARcomp} show box and information sample 
plots for P02 with 95\% confidence level curves from the
least-squares fit.  The information Dimension $D_1$ is shown as
a ``refinement'' of $D_0$, which demonstrates even less bi-scale
behavior, suggesting that the two-slope hypothesis may be an artifact 
of the box-counting method.  The data provides a very clean linear fit
in both cases, generally better for the information dimension $D_1$,
albeit not significantly ($r^2 = 99.9\%~vs~99.8\%$).  Similar behavior 
is observed for the other images.

In fact, the lower-scale measurement process is somewhat 
dependent on the resolution of the image.  Some of the fits suggest
a shallower slope at smaller scale lengths, but it is not necessarily
justified to assume that this behavior is an artifact of the pattern, and
not the resolution of the image, or limitations of the counting/analysis
method.

Although the methods herein and those of 
reference~\cite{taylor} differ interpretationally, roughly
the same end result is obtained.  That is, one can still associate an effective
fractal dimension in the range $D_0 \simeq 1.6-1.8$ with the patterns on
the paintings, simply by considering the slope of the entire fit.
A changing slope from box counting does not immediately imply
multifractal behavior.  

While it may be that
the slope tends to be shallower at lower scales, this may not be an
artifact of the data set so much as a manifestation of estimates and
assumptions about the data.  There may also be lower-level resolution
limitations due to the finite size of the pixels.  Surely, in the mapping
from a 5~metre painting to a 30~cm page -- or to a 1000~px binary image -- there
must be some significant level of information loss at the lower scales
of resolution (in both the photograph and the data scanning process).

There does not seem to be significant variation in dimensions between
lighter and darker colors, although in certain cases it is observed that
the lighter pigment patterns tend to exhibit lower fractal dimensions.
This could be due to a different deposition mechanism than simple
dripping, as well.  It is perhaps a sweeping generalization to assume that
all the pigments were applied in exactly the same fashion.  

So, it becomes somewhat unclear how one can
define the ``dimension'' of the entire image.
This suggests an application of the Fractal Union Theorem (see
{\it e.g.}~\cite{vicsek}).  Since the fractal dimension of the union
of fractals $\bigcup F_i$ has dimension $\max\{D_i\}$, then the 
fractal dimension of the entire image will correspond to that of the 
most complex pattern.  Thus, isolation of the pattern with
the highest dimension can be interpreted to characterize the fractal nature
of the entire image.  This is consistent with the notion of the ``anchor layer''
discussed in references~\cite{taylor} ({\it i.e.} the pattern which seems
to strongly influence the dimension of the whole image).  However, 
note that these authors
mention that the overall dimension {\it increases} as more patterns are
considered, driving the overall dimension to $D \sim 2$.  It is unclear
what is meant by this statement, but from a mathematical approach, it
seems contradictory to the associated theorems.

\section{Robustness of Analysis Method}
\label{robust}
The exact determination of the fractal dimensions depends on the 
cutoff for the colors under consideration.  Thus, there is a certain amount
of variability in the estimation.  To test for further variability (and
hence potential limitations of the box counting method applied to such
images), P01, {\it Reflections of the Big Dipper}, and
{\it Number One 1949} were each rotated by 90$^\circ$, and the corresponding
Box and information Dimensions were calculated for a color radius
of $R_{\rm{RGB}} = 20$ pixels:
\begin{itemize}
\item{{\it Blue Poles}: $D_0 = 1.68 \pm 0.03$; $D_1 = 1.65\pm0.02$}
\item{{\it Reflections ...}: $D_B = 1.77\pm0.04$; $D_I = 1.72\pm0.03$}
\item{{\it Number One 1949}: $D_B =1.73\pm0.05$; $D_I = 1.70\pm0.04$}
\end{itemize}

These are quite commensurate with the values obtained in Table~\ref{poldims}, 
subject to the cited error, confirming the rotational invariance of the result.

Pixels are randomly displaced by 5, 10, and 20 positions from their original
location, and the appropriate dimensions are again calculated for the same
paintings.  Table~\ref{pixd} shows the results for the same paintings as
above.  

\section{Fractals in Gestural Expressionism}
\label{autosection}
If the patterns which appear in these paintings truly are the product
of physical processes, rather than pure artistic expressionism, than such
structure should be visible in similar works by other artists.  Based on
the a similar analysis to that of Reference~\cite{jrmccd}, it seems 
reasonable that other
images formed by similar processes should be classifiable by similar 
statistics.  

Roughly contemporaneous with Jackson Pollock, the Qu\'{e}bec School
{\it Les Automatistes} also produced non--representational art not unlike
the drip paintings studied above.  The group was spearheaded by 
Jean-Paul Riopelle and Marcel Barbeau who collectively produced their works
over the 35~year period spanning 1945-1980.

\begin{figure} \begin{center} \leavevmode
\includegraphics[width=0.5\textwidth]{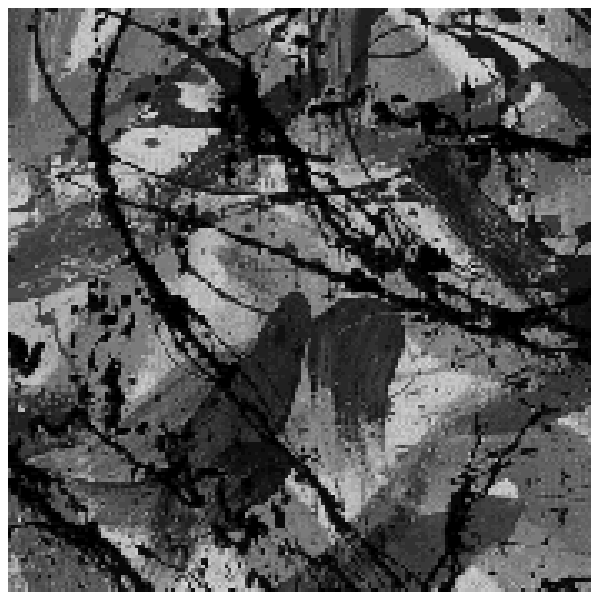}~~
\includegraphics[width=0.5\textwidth]{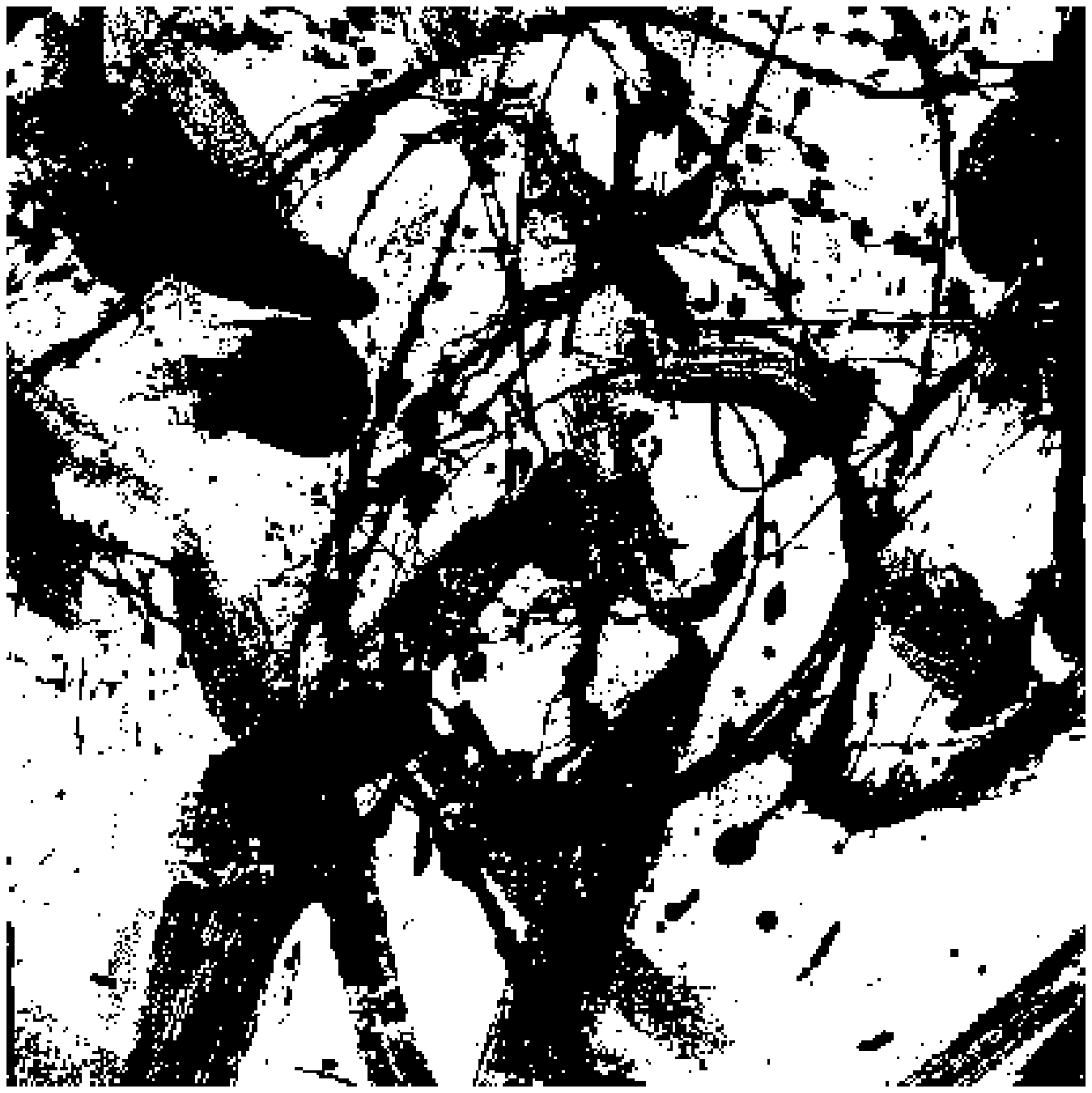}
\end{center} \caption{
Sample of Les Automatiste painting and black pigment pattern (image A05).  }
\label{auto2}
\end{figure}

Figure~\ref{auto2} shows a section of a drip painting from Les Automatistes,
as well as the filtered black pigment pattern.
Table~\ref{autodims} lists the calculated Box and information Dimensions
for select works by {\it Les Automatistes}, subject to the same selection
criteria as before.  

As with Pollock's drip works, the dimensions of the
patterns fall roughly between $1.6-1.8$.  The box and information dimensions do
not explicitly differentiate between Pollock's work and that of Les 
Automatistes.  In fact, the difference in the average fractal dimensions for
each artists was shown to be statistically insignificant using a two-way
ANOVA in Reference~\cite{jrmleo}.  The lighter colors display mildly lower 
dimensions than the darker pigments, although this may be due to cutoff
limitations of the filtering process.  Similar behavior was observed in
the images by Pollock, so whether or not this is an actual artifact of
the pattern or a numerical effect is a subject for future investigations.

While this was somewhat the case with Pollock's works,
there are perhaps sufficient discrepancies to suggest that such a measure could
be indicative of different uses of colors and techniques between these
artists.  This includes using lighter colors for balance in an image,
versus their use for adding contrasting depth.  

In any event,
the general equivalence of the dimensions of Pollock's works and those
of Les Automatistes suggests that the utility of this technique as a 
``fingerprinting'' mechanism for individual artwork/artist association may be 
in vain.  As with the galaxy clustering models, one could assert that the
technique can isolate only construction method, and not structural
variation within
the method.  Those who are dissuaded by the effective reductionist implications
of the analysis may find comfort herein.  In order to further address this
point, the multifractal analysis will be addressed in Section~\ref{multiart}.

\section{Multifractal Spectrum of Non-Representational Images}
\label{multiart}
\begin{figure} \begin{center}
\leavevmode
\includegraphics[angle=270,width=0.75\textwidth]{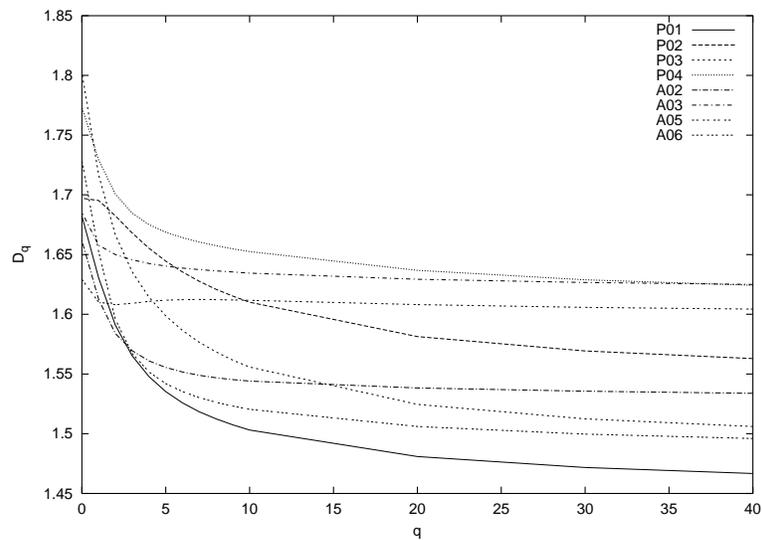}
\end{center} \caption{
Multifractal dimension spectra for select images of Table~\ref{appendix1}.  Erro
r bars are suppressed for easy viewing.
}
\label{MC1}
\end{figure}

Figure~\ref{MC1} shows the range of generalized dimensions
$D_q$ for these patterns.  Note that the overall depth of the generalized
dimension spectrum is not excessive, suggestive that if these patterns
can be described by multifractal statistics, their overall structure
is not that extensive.  Furthermore, note that for the majority of the
cases considered, there is no appreciable difference in the range
or shape of the spectrum.   The errors from the linear fits are generally 
of the order 0.05 or less, but these may be underestimates since 
no error is introduced for variation in the color.
The limiting values of $D_{\infty}$ give
less intuitive insight into the densest clustering regions, unlike in
the case of the three-dimensional sets considered earlier.  

Table~\ref{dqdiffs} shows inhomogeneity measure
for several Pollock and Automatistes works, defined  by Equation~(\ref{diffdq}).
In general, the results suggest that Pollock's works tend to be ``deeper''
than those by the Automatistes ({\it i.e.} greater degree of inhomogeneity), 
perhaps a result of painting styles and refinement techniques.  This
could hint at a potential method of distinction for the sets of similar
classes, but one must be extremely cautious of the selection criteria
for the pattern in question.  It is more likely that these measurements
are simply too ``noisy'' for any useful approximation.

\section{Comparison of Construction Method: Gestural Expressionism {\it versus} Artonomy}
\label{artonomy}
The utility of the analysis methods contained herein seem limited in the
context of analysis of differing works of the abstract expressionist class.  
For the cases considered, the variance in the data seem too small to be of
any particular import for specific identification.
However, when applied to other images, certain differences do arise, enabling
one to make distinctions at least on some level.  

In particular, the
artwork of Tsion Avital \cite{avital} will be considered.
In his seminal work on the subject \cite{avital}, Avital introduces the 
concept of
{\it Artonomy}, the focal blend of artistic expressionism with scientific
order.  A complete description of the intricacies of the method will not
be discussed here, and the interested reader is referred to the aforementioned
citation for further details.  The crucial point is that the construction
``philosophy'' for these images is strictly different than those of the gestural
expressionist class considered previously. 

Avital notes that the concept of Artonomy is based on certain principles of
of ``isotropy'' in the creation process.  There are no preferred sets of
colors, and the use and applications of each color are deemed
``equal'' in value to every other.  Colors (or elements) are combined into
a variety of rigorously-defined mathematical sets (dubbed ``moments''), 
and the final paintings are constructed from combinations of these moments
subject to the appropriate rules.  Paints are applied in a simple manner
({\it e.g.} controlled brush, or ``toothbrush spray'') and as with the color
selection, there is no preferred method.

The moments are methodically positioned on the canvas in a recursive fashion
quite reminiscent of the basic structure of multifractals (such as, for 
example, the framework outlined in Figure~\ref{rboxes}).  Of particular
interest is Avital's ``type $\gamma$'' moment construction rule
\cite{avital}, which operates on the basis of information density on 
the canvas.  Here, he defines the density as low when like colors or
hues are assembled (homogeneous elements), and high with the
neighboring placement of contrasting elements (heterogeneous elements).
Avital defines an {\it abstract field} as one which is comprised of low
density regions, and a {\it concrete field} as one composed of high
density regions.  Abstract and concrete fields may be inter-mixed to form
heterogeneous fields.  Images AV01-03 represent ``homogeneous'' constructs,
while AV04-06 are ``heterogeneous''.

So, in a sense, comparison of gestural expressionist ``structures''
with those of Avital constitutes a contrast in construction 
methodologies -- random versus algorithmic -- and 
thus Avital's works can be taken to be a control or model comparison.

Table~\ref{avdims} shows measured generalized multifractal dimensions for
various color distributions in Avital's works.  It is somewhat difficult
to define an exact base color in the homogeneous images (AV01-03), since
the resulting pattern is due to integrated ``aerosol'' deposition.
In any event, note that unlike the
Pollock and Automatistes images, Avital's works show {\it no
significant multiscaling behavior}.  In many cases, the calculated $D_{\infty}$
is {\it higher} than $D_0$, yielding a negative $\Delta_{0,\infty}$.  
It should be noted that similar behavior was observed
for some monofractals and simple geometric shapes ({\it i.e.} objects for
which there is a single scaling dimension), where the $D_q$ for small $q$
tend to underestimate the actual dimension.  Also, if one considers
the associated statistical error, then these negative values are easily
accounted for.  Thus, it can be concluded that
Avital's systemic blobs are devoid of the ``rich'' structure with which
the gestural expressionists endow their works, due perhaps in part to
the very algorithmic (less random) nature of the construction.  

Furthermore, Avital's homogeneous works ({\it e.g.} AV01) were constructed
from the spray of a paint from a toothbrush.  Thus, the resulting 
structure is probably similar to the deposition from an aerosol source.
The dimensionality most likely reflects this mechanism, to a certain
extent.  Avital's heterogeneous works (AV11, AV12) were constructed
with controlled paint brush strokes.  So, these could actually be
considered two separate sub-construction mechanisms.

%

\section{Reconstructing Images From the Multifractal Spectra}
\label{uses}
Accurate determination of the multifractal spectra of singularities for
a dynamical process can yield important information about its construction
processes and associated constraints.  
As previously mentioned, the $D_q$ provide
important information about the n-tuple ``pair-wise'' clustering behavior
of the set, and provide a unique characterization of the object under
investigation.

The quantities
thus obtained can be used as physical constraints to be used in development
of any model, and can perhaps yield interesting information about
the dynamics of the pattern generator during the construction phase.  
Since the multifractal analysis herein seems only to have the ability
to discern one class of structure from another, one must ask whether
or not there is a useful tool to distinguish between like sets.  A short
analysis is performed herein on the like image arrays of the
abstract expressionist class, in order to address this problem.

\begin{figure} \begin{center}
\leavevmode
\includegraphics[angle=0,width=0.8\textwidth]{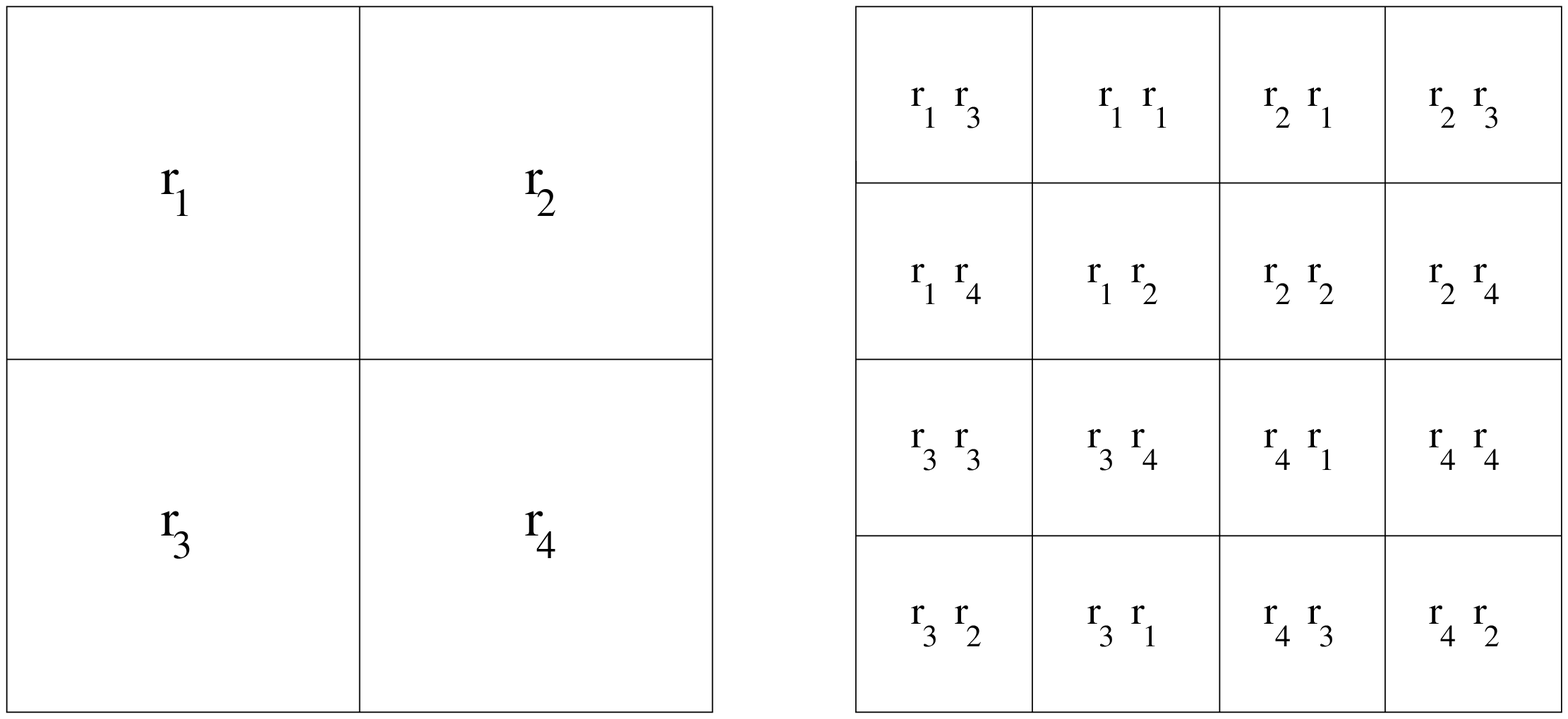}
\end{center} \caption{
Multifractal scaling behavior, showing 1-level reduction of distribution
with probabilities $r_1, r_2, r_3,$ and $r_4$.
}
\label{rboxes}
\end{figure}

By definition, a multifractal is an inhomogeneous recursive scaling (a
multifractal lattice).  Suppose a square (or box, to be consistent with
the current nomenclature) is divided into four sub-units of equal area.
Then, one can describe the relative portion of the pattern contained in
each box by the probabilities $r_1, r_2, r_3$, and $r_4$ respectively
(see Figure~\ref{rboxes}). 

At the next level of recursion, the weights $r_i$ are {\it randomly} reassigned
to each sub-box of the previous layer, and the process repeats (cascades)
down to any level of recursion desired.

Recall that the generalized dimensions are calculated from the 
partition function (\ref{zpart}), and furthermore $(q-1)D_q = \tau(q)$.
From (\ref{taudef}), one can estimate the difference between two
successive cut scales $\delta$ and $\delta/2$ ({\it c.f.} Figure~\ref{rboxes})
as
\beq
D_q(q-1) = \tau(q) = \frac{\Delta\log[Z(q,\delta)]}{\Delta\log[\delta]}~,
\eeq
In terms of the probability $r_i$ for each box, this becomes
\beq
\frac{\Delta\log[Z(q,\delta)]}{\Delta\log[\delta]}
= \frac{\log[\sum_i r_i^q] - \log[\sum_j r_j^q \sum_k r_k^q]}{\log[\delta] 
- \log[\delta/2]}~,
\eeq
which reduces to
\beq
D_q(1-q) = \frac{\log[Z(q,\delta)]}{\log(2)}~,
\eeq
So, one can substitute $Z(q,\delta) = r_1^q +r_2^q+r_3^q+r_4^q$ to
obtain
\beq
r_1^q +r_2^q+r_3^q+r_4^q = 2^{D_q(1-q)}~,
\label{simpdq}
\eeq
and the distribution probabilities $r_i$ may be obtained from a system of
four equations.  Note that this expression may be simplified, by
noting the constraint $r_1 + r_2 + r_3 + r_4 = 1$.  Furthermore, the
$q=2$ version of Equation~\ref{simpdq} represents the equation of a
4-sphere, whose roots may be easily obtained.

Hence, the system of four equations may be reduced to a system of two
unknowns, in this case $r_3$ and $r_4$.  The values of the four possible
$r_i$ may be isolated by optimizing the possible values of $r_3, r_4$ which fit the measured $D_q$ spectrum of generalized dimensions.  This is achieved
by finding sets of $r_i$ for which the individual separations 
$\Delta D_i = D_{\rm calculated} - D_{\rm measured} < \epsilon$, for
$\epsilon \sim 0.0001$ on average.

In Table~\ref{shapers}, the $r_i$ values for various shapes of known monofractal
dimension are presented.  Note that for a figure of topological dimension
$D_T = 1$ ({\it e.g.} the line), the weighting factors suggest that
for the appropriate cut of the plane in Figure~\ref{rboxes}, the shape will
only have a nonzero probability of being in any two of the 4 sub-boxes,
a result which certainly makes sense.  Similarly, a figure of dimension
$D_T = 2$, which ``fills the plane'', will have equal probability of being
in every box.  The negative component for the Koch Curve (Island) is 
most likely the result of numerical uncertainty, since negative probabilities
would not make sense.
The calculated $r_i$s for the Koch Curve and Sierpinski Gasket can also be
interpreted to reflect the construction algorithms and symmetries for 
each figure.  

Table~\ref{dqrs1} lists the calculated values of $r_i$ for 
the ``anchor layer'' pigment shapes in several of the images considered 
previously.  The values are relatively consistent for each painting,
although this is not a particularly surprising result, since the $D_q$
spectra themselves are not significantly different.  

Each set is characterized by a rather even distribution amongst three
of the boxes, and a fourth which is smaller by an order of magnitude.
The almost homogeneous distribution is no doubt reflective of the fact that the
generalized dimensions are close to 2.  It can be shown that for a 
distribution with $r_i = 0.25$ for all $i$, the generalized dimensions
all collapse to 2 (or vice versa).

It may be somewhat discouraging to note that these values are rather close
to one another, and are seemingly indistinguishable.  However, it should
be noted that the formalism outlined above is not a singular representation
of a multifractal scaling process.  Four quadrants have been used to show
recursive scaling in part for computational efficiency.  This could be
a significant source of error if the scaling behavior is radically different
than this model requires.  Additionally, this may again be
a fundamental problem with the resolution limitations of the method.

The parameters herein can conceivably be used in the formulation of 
a physical model which could reproduce the associated images, at least
on a statistical level.  Furthermore,
the authors of \cite{taylor1} have studied video recordings of Jackson
Pollock in his creative process, and have found that the ``fractality''
of the overall work took less than a minute to define.  Surely, this provides
an additional constraint on a such a cascading model.  

On a subjective level, one wonders whether or not
the smaller fourth quadrant could conceivably be interpreted as an imprint
of the presence of the ``source'' of the image pattern (in this case, the
physical presence of the artist).  That is, at any point during
the construction of the painting, the artist has free choice to paint in
three of the four ``quadrants'' (the last being occupied by himself).
Thus, this could be nested in the recursion, and detectable by such an
analysis.  If this explanation were to accurately represent the evolution
of the pattern, it could be used to distinguish between patterns 
constructed by humans, and those created by machines or other natural
processes.

\section{Visual Multifractals}
\label{visual}
The analysis in the previous Sections relied predominantly on a 
color filtering process dependent on the distance in RGB space
of pixel color to its target ``match''.  However, many reports suggest
that the hierarchical clustering of the images has some variety of
psychological effect on the viewer.  While using RGB primaries as the
filtering criteria isolated the {\it physical} structure of the 
blob, it may {\it not} be an effective measure of the {\it perceived}
structure.  Taylor {\it et al.} have recently studied physiological
responses to fractal viewing, and have concluded that observers do 
exhibit definite responses when presented with certain fractal patterns
\cite{taylor2,taylor3}.

The problem of structure identification and discrimination is not a new 
one in psychological
circles, nor is it by any means a solved one.  Implicitly related to this
topic, the authors of reference~\cite{gerry1}
discuss the perceptability of hierarchical structures in abstract or
non-representational constructs (whose subject matter is used in a 
comparative study in Section~\ref{artonomy}).  
In fact, rapid object
recognition and categorization via boundary isolation versus ``blob''
identification is a subject of growing scientific interest (see
\cite{schyns2} and related references therein).

A complete understanding of the nature of color perception is still lacking.
Thus, the notion of a {\it visual fractal} is introduced
in contrast to those fractals previously considered.   Instead of
direct observation of colors, the focus is instead shifted to
{\it edge structures}.  
This is effectively an analysis of luminance gradients
within the image, and not directly on the RGB color field distribution
(although the luminance values are determined by R, G, and B mixes).
In fact, after completing this research, the work of 
references~\cite{turiel} was discovered.  Therein,
the authors discuss the potential uses of measuring the multifractal spectrum 
of luminance gradients in natural color images, to determine whether
or not it conveys relevant information about the image.  The analysis
presented herein is quite similar in these respects, and thus is not
performed without physical justification.

\subsection{Luminance Edges as Visual Fractals}
\label{luminance}
While ripe with theory, the actual dynamics of human color visual 
processing are poorly understood,
yet it is clear that one does not {\it require} a wealth of color information
to visualize a scene.
A subject of ongoing interest (see {\it e.g.} \cite{schyns} and similar
references) is whether or not object/pattern recognition occurs on the
level of ``blob'' or ``edge'' identification.  Studies of eye moments
in subjects viewing artistic scenes seem to support the notion that
human fixate on particular aspects of an image, supporting the notion
that ``blobs'' are viewed \cite{yarbus} but it is perhaps unclear as to
how these objects are distinguished.  

Similarly, the images formed by one's brain may {\it not} be fully representative
of the scene which one views.  Both chromatic and achromatic information
received from stimulation of the photopigment receptors in the rods and
cones, are ``preprocessed'' before being sent to the visual cortex via
the optic nerve.  

In a similar vein, it is useful to find a ``one-parameter'' method
of analysis for such color images, as an attempt to find a suitable way
to discriminate between them.  The results of the previous Sections
suggest that different choice of colors yield somewhat differing dimensions,
so it would be helpful to find an element common to all images which 
is independent of any particular color.  Thus, one can consider analyzing
{\it luminance} properties of the image.

Edge detection in the visual system occurs on several different levels, 
although it is not necessarily know which one is ``dominant''.
One such mechanism is known as {\it lateral inhibition} (LI).
In short, this process measures the relative excitatory signal output from one
photoreceptor with inhibitory signals from adjacent neighbors, effectively
producing a difference output signal which is sent to the visual cortex.
The result is that the strongest excitatory signals will be sent from
those retinal neurons which detect {\it luminance changes} across the
field \cite{hcv}.

\begin{figure} \begin{center}
\leavevmode
\includegraphics[width=0.3\textwidth]{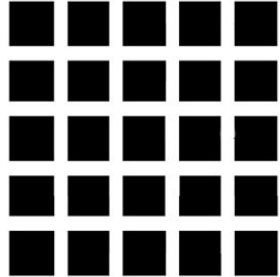}
\end{center} \caption{
Black and white checkerboard pattern (``Hermann Grid'')
showing visual luminance structure variation due
to edge enhancement effects in the visual processing system.
}
\label{checkbrd}
\end{figure}

Coupled to the visual system's ability to interpolate information in a
field from missing stimuli ({\it e.g.} as with the blind spot), LI
can create artificial luminance and brightness variation
effects which are not physically present in the original scene \cite{hcv}.  
For example,
a black and white checkerboard will seem to have greyscale variations
across the pattern.  The intensity (luminance) of the central 
squares is the same in each case, but the square surrounded by white 
appears to be darker than the other (see Figure~\ref{checkbrd}).  This
exemplifies the eye's ability to create artificial variations in scenes
which are otherwise not physically present.  Hence, this provides a rather
simple example of how visual interpretation
of an object may not be complete commensurate with the actual physical
characteristics.

Lateral inhibition is, however, only one of several mechanisms responsible
for the detection of contrast edges in a visual field.  While LI mechanisms
operate in the eye, such detection is known to occur in the visual cortex
itself.  Hubel and Wiesel were responsible for the discovery of ``orientation
columns'' within the visual cortex, cells responsible for 
the identification of specific edges or boundaries orientations.  The
aforementioned researchers share the 1981 Nobel Prize in Medicine for their
research efforts.  The interested reader is directed to reference~\cite{hubweis}
for an expository account of their work.
Thus, there is sufficient physiological and psychological motivation to
consider possible structural differences in contrast edges.

The transformation
from RGB primaries is of the form $Y = 0.299\;R + 0.587\;G + 0.114\;B$
\cite{graphics} (note that the color coordinates must be normalized), 
which implies pure white coordinates $(R,G,B) = (1.0, 1.0, 1.0)$.
Note the relatively higher weighting of R and
G primaries to B.  This is reflective of the eye's sensitivity to similar
wavelength intensities.  
In fact, these roughly correspond to the three
basic types of cone cells with similar thresholds, denoted as L, M, and
S (for long, medium, and short wavelengths).  
This is actually one component of a separate CIE color system known as
YIQ (the channels I and Q are encode chromacy information, hue and saturation).
The luminance channel is what one generally associates with
greyscale images, an in fact is that information which is transmitted
in black and white television signals \cite{graphics}. 
Edge detection is performed by generally-available image manipulation tools,
which measure the vector sum of two perpendicular Sobel gradient 
operators (see {\it e.g.} \cite{sobel} for more information).

These are perhaps crude approximations to the actual physiological processes
at hand, so implicit limitations in the estimates should be accordingly
recognized.  Certainly, the method does not purport to be a realistic model
of the visual system.
It should, however, provide a decent first-pass approximation
to any inherent structures and effects therein.

\subsection{Pollock {\it vs} Les Automatistes}
\begin{figure} \begin{center}
\leavevmode
\includegraphics[width=0.5\textwidth]{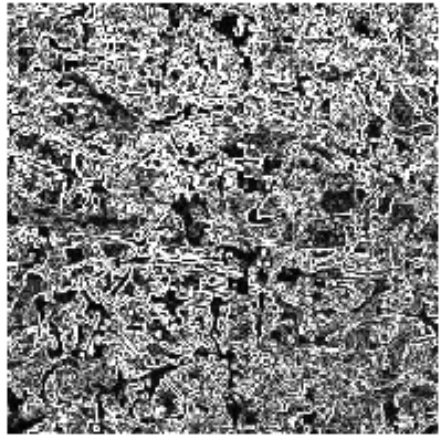} 
\vskip 0.5cm
\includegraphics[width=0.5\textwidth]{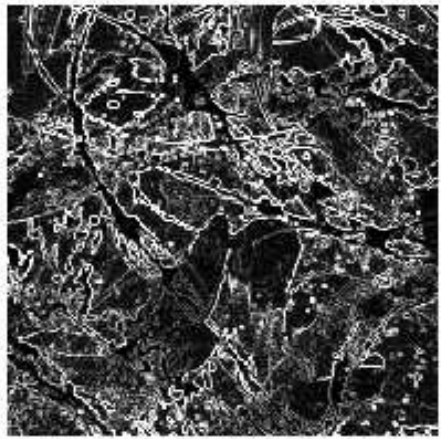}
\end{center} \caption{
Sample luminance gradients for P02 (left) and A05 (right).
}
\label{p32E}
\end{figure}

\begin{figure} \begin{center} \leavevmode
\includegraphics[angle=270,width=0.75\textwidth]{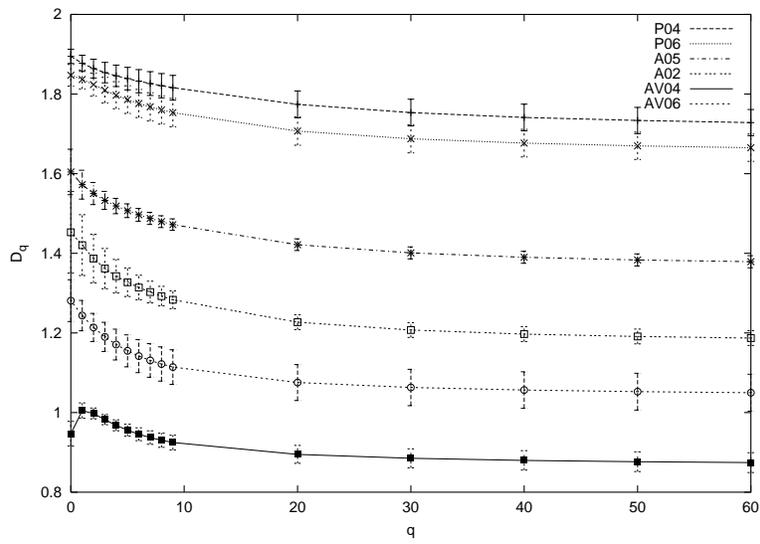}
\end{center} \caption{
$D_q$ spectra for luminance gradients of Pollock, Automatistes, and
Avital images $(\beta=1)$.  The structures are clearly separated by
$D_0$ values, although the overall structure of the curve is similar.
}
\label{p32Efa}
\end{figure}

Figure~\ref{p32E} shows a sample edge-detect transform for a images of
Pollock and Les Automatistes, with the associated
$D_q$ spectra in Figure~\ref{p32Efa}.  For the color-filter process,
the target color in
this case is pure white, and the color 
radius is taken to be the linear
distance away from the point.  Thus, for a small radius, the images
with the highest gradients will have the largest dimensions.  
Table~\ref{polautodims} and \ref{avedges} give dimensions for
both $\beta=1$ and $\beta=30$, which give an indication of the ``value'' of
the strongest gradients.  
It should be noted that since neuronal firings are 
triggered by threshold-breaking stimuli, a discrete 
cutoff is more realistic than a Gaussian drop-off.

The calculated box dimensions for the edge-detected Pollock images tend to be 
higher than for the individual blobs, generally $D_0 > 1.8$.  This can
be interpreted as implying that the luminance edges form a much more
complex visual field, and that the edge lines are more ``space filling'',
and providing a ``busier'' or ``fuller'' visual experience.

When the method is applied to the gestural expressionist 
works of Les Automatistes, differences become more apparent
(see Figure~\ref{p32Efa}). 
The box dimensions for Les Automatistes is generally
lower (albeit not much) 
than those of Pollock's.   Similar results are obtained for
other images (see Table~\ref{polautodims}).  Note that the measured dimensions
do not increase significantly from $\beta=1$ to $\beta=30$. 

This suggests that there is a potential visual difference between images
by these different artists.  While the final products may resemble each
other at first glance, the intricacies of the two images from a luminance
gradient / visual standpoint appear quite different.  Again, in Reference~\cite{jrmleo}
the difference in average fractal dimension of the ``edge'' patterns was 
determined to be statistically significant.

Based on the work of previous authors \cite{pickover} and their own survey 
on preferential response to drip patterns, the authors of \cite{taylor1} 
conclude that patterns possessing a fractal dimension of roughly 1.8 are 
inherently aesthetically-pleasing to the observer.  A follow-up study
suggests that ``creative individuals'' have a preference for high values
of $D$ \cite{aks}.   One could imagine that this type of perceived structural
difference could contribute to an observer's ``appreciation'' of one image
or style over another.  

\subsection{Comparison With Avital}
Avital's definitions of homogeneous and heterogeneous fields (not to
be confused with homogeneous fractal distributions), along with the
concept of information content, are a natural extension of the 
notions of luminance gradient structures proposed in Section~\ref{luminance}.
In fact, the very notion of information content is at the heart of 
the multifractal formalism.  Thus, a luminance-gradient 
analysis of Avital's images should reveal certain properties about the
formulaic construction of the pieces, or at the very least lend contrast
to the more psychological algorithms used by the Abstract Expressionist
artists (or perhaps any other artist). 

\begin{figure} \begin{center} \leavevmode
\includegraphics[angle=90,width=0.7\textwidth]{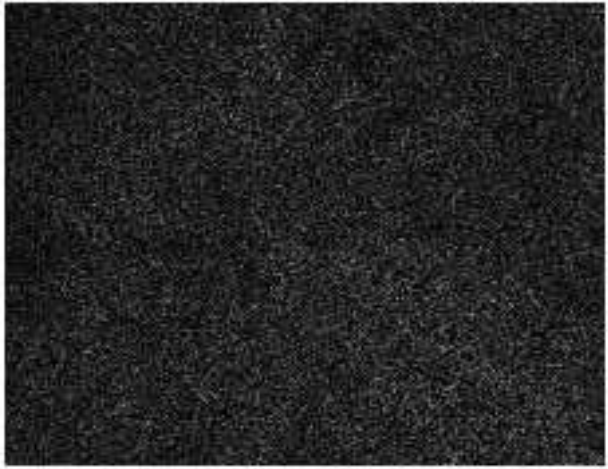}
\vskip 1cm
\includegraphics[angle=90,width=0.7\textwidth]{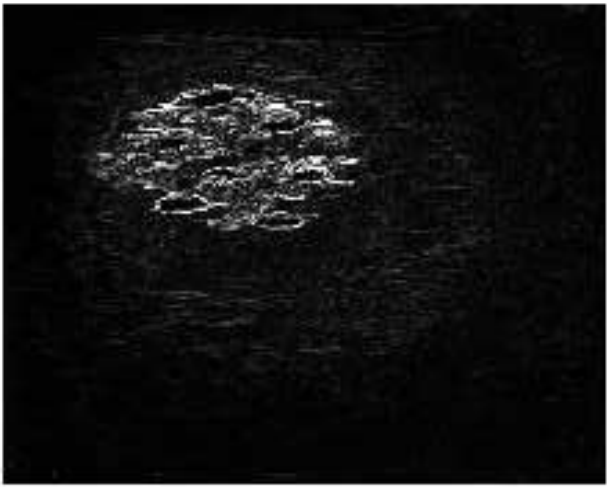}
\end{center} \caption{
Sample luminance gradients of images for AV01 (top) and AV06 (bottom).
}
\label{avedge}
\end{figure}

Table~\ref{avedges} shows the effective edge-detection dimensions of
various works by Avital \cite{avital}, as well as a rough definition of the
type of image.  Figure~\ref{p32Efa} shows the associated spectra, in
comparison to the previous images.
Since reproduction of every image in this work is not
warranted, The previous color panels demonstrate the general qualities of
each type of image (labeled homogeneous and heterogeneous), while
Figure~\ref{avedge} shows the resulting luminance gradients.  
It should be noted that the images classified as ``homogeneous'' all
conform to Avital's ``S/D/$\delta$'' construction algorithm (made from
the spray of a toothbrush) \cite{avital},
which indeed embodies an inherently smooth transition to complexity vis-a-vis
color selection and application.  In contrast, the images noted as
``heterogeneous'' are from Avital's ``S/C/$\gamma$'' algorithm,
which allows for a counterbalance between abstractness and concreteness.
These are simply painted with controlled brush strokes.

The lower dimensions and higher error for the homogeneous images 
in Table~\ref{avedges} demonstrate the 
low color contrast nature of the images, and hence the shallow depth of
luminance variation across the canvas.   
In particular, note that while
image AV01 is physically a mix of bright pigments, 
there is virtually no strong luminance gradient 
across the canvas (hence to effective dimension of 0).  
These dimensions imply there
is little ``luminance information'' in the fields.  
As the images begin to approach
heterogeneity, the background field is contrasted with patches of color,
whose overall boundaries are quite regular.  This is indicative by the
relatively low range of $D_0 = 1.1-1.3$, in contrast to the exceedingly
high dimensionality of the gestural expressionist paintings, as seen
in Table~\ref{polautodims}.  The latter is indicative that the luminance
gradients densely fill the canvas for this particular school or movement,
while Avital's shapes are more concentrated and simple. 

Thus, Avital's homogeneous work is less ``interesting'' from an
edge detect view than the Pollock or Les Automatistes images considered
(there is less edge information conveyed about the scene),
while the heterogeneous work brings focus to 
particular objects via these edges (although still with much lower
dimension that the gestural expressionist images).

These results suggest that this analysis method could
distinguish between sources, as well as construction
mechanisms of the images.  Further investigation would be required.
However, whether or not this type of distinction is possible, 
identification of such structural signatures could have applications
in external fields.
For example, albeit beyond the scope of physics, visual detection
of fractal structure in luminance gradients could have profound
consequences for the fields of aesthetics and visual appreciation
of complex scenes ({\it e.g.} what qualities makes an image interesting to
us?).

It is interesting to note that Avital himself classified works such as 
Pollock's as ``moment type $\omega$'', whose paradigm rests on the notion of
``arbitrariness'', in which the combinations of elements (moments) are
scattered at personal will about the canvas (he further notes these to
require ``minimal capacity of inventiveness'', and casts Pollock's
art as containing nothing meaningful or interesting \cite{avital}).
Avital is careful to note the distinction between ``arbitrariness''
(whose choice of elements is human) and ``randomness'' (whose source is 
instead probabilistic).  Unfortunately, Avital presents no simulations of
type-$\omega$, so it is not possible to compare these with the works of
the gestural expressionists considered previously.

\section{Potential Limitations of the Method}
\label{limitations}
Of course, the method described herein is not without limitations, and
is only designed to be a ``first-order'' attack of the issues at hand.
As discussed previously, digital image analysis techniques provide a 
statistical description of the {\it entire physical image}, with no
regard for perceptual interpretations by observers.  The calculated
dimensions assumed equal weighting for all portions of the canvas, when
in fact (depending on the distance from which they view the scene)
observers will not register all portions of the field equally.
Both rod and cone cells are unevenly distributed about the retina, with
a disproportionately large number of cones clustered in the fovea
centralis \cite{hcv}.  This cone clustering is crucial for perception
of color and fine visual detail via fixation, and is the primary reason for
the drop in acuity in peripheral vision.  

So, if the image of interest
fills the visual field, only the central-most regions will convey the
largest amount of information.  However, this should not necessarily
affect the overall ``visual estimation'' of the fractal nature of the
piece, although edges may become more blurred (resulting in potential
shifts in a ``visual multifractal spectrum'').
Furthermore, the method does not account for other biasing effects such
as color blindness, or any visual acuity drops ({\it e.g.} myopia or
other focal abnormalities).  The robustness tests of 
Section~\ref{robust} suggest that the dimensionality of patterns will
increase for dispersive patterns (as they should, approaching homogeneity),
which could replicate such vision problems.

Finally, the methods outlined herein do not all complete correspond to 
actual physiological processes which occur in the eye.  Reference~\cite{hcv}
provides several alternative color space transformations which are perhaps
more appropriate for the actual analysis of cone/photoreceptor
excitations from lightness/luminance and chromatic stimuli.  A full 
investigation and implementation of these methods is discussed in
Reference~\cite{jrmcie}.

\section{Concluding Remarks and Future Directions}
\label{conclusions}
The use of fractal and multifractal analysis as a discriminator or 
fingerprint method for classifying abstract expressionist art is a budding 
field.  However, the available results are indicative that the method may
well yield promising results.  The fractal signatures obtained from paint
blobs are not significantly different from one another, implying that this
method is not useful for ``authenticating'' works by any one particular
artist within a movement.  It apparently does differentiate between the
movements themselves.  This is similar to the behavior observed in
\cite{jrmccd}, where the multifractal spectrum was shown to differentiate
between galaxy cluster formation mechanisms, but could not discriminate 
between instances within the same model.  
The ``edge multifractal'' does yield differentiable results, curiously,
which based on aspects of visual processing lends to the interpretation that 
this could represent some type of ``aesthetic preference''.

One of the motivational questions which inspired the fractal analysis
of gestural expressionist art is: ``Does there exist an inherent structure
within the painted patterns which one perceives, and hence yields 
an unconscious psychological effect on the observer?''.
Rephrased, on can pose the question: does the brain possess a mechanism
whereby the observer can gain information from a scene previously unknown
to them?

This question certainly addresses the very heart of recognition and learning
methodologies, but unfortunately the exact neural mechanisms which lead to
cognition are not well understood.  Recent discoveries in Neuroscience
have paved the way for a potential revolution in this field, however.

Recent studies have revealed striking neural activity in several species
of primates which respond not only to physical imitation of observed
movements by others, but also {\it passive observation} of such actions.
That is, such neural firings are indicative that the individual need
not repeat the action in order to cognitively process its meaning --
quite literally, a case of ``monkey-see, monkey-do''.  Based on these
imitation characteristics, such cells have been dubbed {\it mirror neurons}.
For a basic introduction, see \cite{mirror1} and references therein.

Additional studies suggest that mirror neurons may be present in higher
species of primates (and in particular may be central to the development
of language skills in humans \cite{rizoarbib}).  If observation of action 
can trigger their firings and initiate comprehension of its meaning, then
it may not be unreasonable to expect that observation of the {\it trace
of an action} can also prompt similar neurophysiological responses.

The authors of \cite{taylor1} note that many natural patterns possess
multifractal scaling behavior, but these are not ``art'' per se.  What
is the underlying differentiator, then, that ascribes to these 
statistically-similar patterns the label of ``art''?
Thus, by observing a complex but statistically-ordered scene such as
Pollock's art, mirror neurons could help to bridge the gap between 
the initial visual processing and associative comprehension and
appreciation of the {\it actions} required to form the work \cite{arbib2}.  
Further study into these hypotheses are currently underway.

\vskip 3cm
\noindent{\bf Acknowledgments}\\
We gratefully thank Tsion Avital for permission to reproduce his art.
This work made possible by grants from the Natural Sciences and Engineering
Council of Canada (NSERC) and by financial support from the Walter C.\
Sumner foundation.

\begin{table}
\begin{center}
{\begin{tabular}{c l c}\hline
ID & Title (Date) & Dimensions (cm$^2$) \\ \hline
P01 & {\it Blue Poles} (1952) & 486.8 $\times$ 210 \\
P02 & {\it Autumn Rhythm} (1950)& 525.8 $\times$ 266.7 \\
P03 & {\it Lavender Mist} (1950)& 300 $\times$ 221 \\
P04 & {\it Reflections of the Big Dipper} (1947) & 111 $\times$ 92.1 \\
P05 & {\it Number One A 1948}   & $172.7\times$ 264.2\\
P06 & {\it Number One 1949}  &  160 $\times$ 259 \\ \hline
A01  & {\it Au chateau d'Argol} (1947) & 55 $\times$ 50.5 \\
A02  &{\it Fi\`{e}vres}  (1976) & 65 $\times$ 80\\
A03  &{\it Tumulte} (1973) & 81 $\times$ 101.5\\
A04  &{\it Voyage au bout du vent} (1978) & 137 $\times$ 188\\
A05  &{\it Suite marocaine no.\ 1} (1975) & 81.3 $\times$ 101\\
A06  & {\it La danse et l'espoir} (1975) & 81.5 $\times$ 101 \\
A07  & {\it Sans titre, Montr\'{e}al} (1959) & 56 $\times$ 43 \\ \hline
\end{tabular}}
\end{center}
\caption{
Catalog of Jackson Pollock (P01-06) and Les Automatistes (A01-07) 
images used in analysis.
}
\label{appendix1}
\end{table}

\pagebreak

\begin{table}
\begin{center}
{\begin{tabular}{c c c c}\hline
Fit& $\beta =$~10 & 20 & 30 \\ \hline
\multicolumn{4}{l}{P01 (black)} \\
$D_0$& 1.51 (0.05) & 1.68 (0.03) & 1.72 (0.03) \\
$D_1$& 1.46 (0.02) & 1.63 (0.02) & 1.67 (0.02) \\ \hline
\multicolumn{4}{l}{P01 (red)} \\
$D_0$& 1.33 (0.07) & 1.54 (0.05) & 1.64 (0.04) \\
$D_1$& 1.22 (0.03) & 1.42 (0.02) & 1.54 (0.02) \\ \hline
\multicolumn{4}{l}{P02 (black)} \\
$D_0$& 1.66 (0.03) & 1.70 (0.03) & 1.72 (0.03) \\
$D_1$& 1.66 (0.02) & 1.70 (0.02) & 1.72 (0.02)\\ \hline
\multicolumn{4}{l}{P03 (black)} \\
$D_0$& 1.73 (0.06)& 1.80 (0.05)& 1.84 (0.04) \\ 
$D_1$& 1.64 (0.05)& 1.72 (0.04)& 1.76 (0.03) \\ \hline
\multicolumn{4}{l}{P04 (black)} \\
$D_0$& 1.70 (0.05)& 1.77 (0.04)& 1.81 (0.04) \\
$D_1$& 1.67 (0.03)& 1.73 (0.03)& 1.77 (0.03) \\ \hline
\multicolumn{4}{l}{P05 (black)} \\
$D_0$& 1.72 (0.04)& 1.77 (0.04)& 1.80 (0.03) \\
$D_1$& 1.65 (0.03)& 1.70 (0.02)& 1.74 (0.02) \\ \hline
\multicolumn{4}{l}{P06 (blue-grey)} \\
$D_0$& 1.64 (0.06)& 1.73 (0.05)& 1.78 (0.04) \\
$D_1$& 1.60 (0.05)& 1.68 (0.04)& 1.73 (0.03) \\ \hline
\multicolumn{4}{l}{P06 (cream)} \\
$D_0$& 1.52 (0.10)& 1.71 (0.06)& 1.76 (0.04) \\
$D_1$& 1.49 (0.08)& 1.65 (0.05)& 1.70 (0.04) \\ \hline
\end{tabular}}
\end{center}
\caption{
Box ($D_0$) and information ($D_1$) estimates for select Jackson
Pollock paintings.  
}
\label{poldims}
\end{table}

\pagebreak

\begin{table}
\begin{center}
{\begin{tabular}{c c c c}\hline
Painting & $\beta$ = 5 & 10 & 20 \\ \hline
Blue Poles & 1.70 (0.03) & 1.72 (0.02) & 1.75  (0.02) \\
 & 1.65  (0.02) & 1.67 (0.02) & 1.71  (0.02) \\
Reflections&1.81  (0.03) & 1.84 (0.03) & 1.87  (0.02) \\
 & 1.76 (0.02) & 1.79 (0.01) & 1.84 (0.01) \\ \hline
Number One 1949& 1.76 (0.04) & 1.79 (0.03) & 1.82 (0.03) \\
 & 1.72 (0.02) & 1.75 (0.02) & 1.80 (0.01) \\ \hline
\end{tabular}}
\end{center}
\caption{
$D_B$ (top row) and $D_I$ (bottom row) measurements for random pixel 
displacements of 5, 10, and 20 pixels.
Error (in brackets) is that of the least-squares fit.
}
\label{pixd}
\end{table}

\pagebreak

\begin{table}
\begin{center}
{\begin{tabular}{l c c}\hline
Image &  $D_0$ &  $D_1$ \\ \hline
A01 (black) & 1.92 (0.01) & 1.88 (0.01) \\ 
A02 (black) & 1.66 (0.05) & 1.61 (0.03) \\ 
A02 (blue) & 1.67 (0.07) & 1.60 (0.01) \\
A03 (black) & 1.69 (0.03) & 1.66 (0.01) \\
A03 (blue) & 1.61 (0.06) & 1.56 (0.04) \\
A04 (black) & 1.57 (0.07) & 1.54 (0.05) \\
A04 (blue) & 1.61 (0.07) & 1.53 (0.04) \\ 
A05 (black) & 1.63 (0.03) & 1.61 (0.02) \\ 
A05 (green) & 1.57 (0.05) & 1.47 (0.02) \\
A06 (black) & 1.73 (0.05) & 1.65 (0.02) \\ \hline 
\end{tabular}}
\end{center}
\caption{
$D_0$ and $D_1$ for various Les Automatistes images, $\beta = 20$.
}
\label{autodims}
\end{table}

\pagebreak

\begin{table}[h]
\begin{center}
{\begin{tabular}{l c c c}\hline
Painting & $D_0$ & $D_{\infty}$ & $\Delta D_{0,\infty}$ 
\\ \hline
P01 & 1.68 (0.03) & 1.45 (0.05) & 0.23 (0.06) \\
P02 & 1.70 (0.03) & 1.54 (0.04) & 0.16 (0.05) \\
P03 & 1.80 (0.05) & 1.47 (0.03) & 0.33 (0.06) \\
P04 & 1.77 (0.04) & 1.60 (0.04) & 0.17 (0.06) \\
P05 & 1.77 (0.03) & 1.55 (0.05) & 0.22 (0.06) \\ \hline
A01 & 1.92 (0.01) & 1.85 (0.03) & 0.07 (0.03) \\
A02 & 1.66 (0.05) & 1.53 (0.05) & 0.13 (0.07) \\
A03 & 1.68 (0.03) & 1.62 (0.05) & 0.06 (0.06) \\
A04 & 1.57 (0.07) & 1.34 (0.05) & 0.13 (0.09) \\
A05 & 1.63 (0.03) & 1.60 (0.07) & 0.03 (0.08) \\ \hline
\end{tabular}}
\end{center}
\caption{
Inhomogeneity measure $\Delta D_{0,\infty} = D_0 - D_{\infty}$ comparison
between Jackson Pollock paintings and Les Automatistes works for anchor
layers ($\beta = 20$).   
}
\label{dqdiffs}
\end{table}

\pagebreak

\begin{table}[h]
\begin{center}
{\begin{tabular}{l c c c l}\hline
Painting & $D_0$ & $D_{\infty}$ & $\Delta_{0,\infty}$ & Color \\ \hline
AV01 (XI) & 1.56 (0.03) & 1.51 (0.08) & 0.05 (0.09) & Grey-green   \\
AV02 (XIII) & 1.68 (0.03) & 1.71 (0.05) & -0.03 (0.06) & Yellow \\
AV03 (XV) & 1.61 (0.02) & 1.60 (0.09) & 0.01 (0.09) & Blue \\
AV04 (III) & 1.45 (0.03) & 1.38 (0.07) & 0.07 (0.08) & Red \\
AV05 (VII) & 1.71 (0.02) & 1.82 (0.07) & -0.09 (0.07) & Black \\
AV06 (VIII) & 1.57 (0.04) & 1.43 (0.09)& 0.14 (0.10)& Yellow \\ \hline
\end{tabular}}
\end{center}
\caption{
Generalized dimensions for selected Avital images.  Image sources
are identified by Plate number from \cite{avital}.  
}
\label{avdims}
\end{table}

\pagebreak

\begin{table}[h]
\begin{center}
{\begin{tabular}{l|c||c|c|c|c}\hline
Shape & $D_F$ & $r_1$ & $r_2$ & $r_3$ & $r_4$ \\ \hline
Line & 1.00 & 0.00 & 0.50 & 0.00 & 0.50 \\
Plane & 2.00 & 0.25 & 0.25 & 0.25 & 0.25 \\
Koch Island & 1.26 & -0.05 & 0.44 & 0.18 & 0.43 \\
Sierpinski Gasket & 1.57 & 0.00 & 0.34 & 0.32 & 0.34 \\ \hline
\end{tabular}}
\end{center}
\caption{
Distribution probabilities $r_i$ for various Euclidean shapes and
geometric fractals.
}
\label{shapers}
\end{table}

\pagebreak

\begin{table}[h]
\begin{center}
{\begin{tabular}{l|c|c|c|c}\hline
Painting & $p_1$ & $p_2$ & $p_3$ & $p_4$ \\ \hline
P01 & 0.03 & 0.27 & 0.30 & 0.40 \\ 
P02 & 0.04 & 0.32 & 0.32 & 0.32 \\
P04 & 0.06 & 0.32 & 0.29 & 0.33 \\ 
P05 & 0.03 & 0.34 & 0.29 & 0.34 \\
P06 & 0.03 & 0.30 & 0.29 & 0.38 \\ \hline
A01 & 0.12 & 0.30 & 0.29 & 0.29 \\
A02 & 0.02 & 0.38 & 0.26 & 0.34 \\
A03 & 0.03 & 0.33 & 0.32 & 0.32 \\
A04 & 0.01 & 0.41 & 0.36 & 0.22 \\ 
A05 & 0.01 & 0.33 & 0.33 & 0.33 \\ \hline
\end{tabular}}
\end{center}
\caption{
Select distribution probabilities $r_i$ for various Jackson Pollock
and Les Automatistes works (anchor layers).
}
\label{dqrs1}
\end{table}

\pagebreak

\begin{table}[h]
\begin{center}
{\begin{tabular}{l c c}\hline
Painting & $D_0 (\beta=1)$ & $\beta=30$  \\ \hline
P01 & 1.76 (0.03) & 1.78 (0.03)\\
P02 & 1.90 (0.02) & 1.90 (0.02)  \\
P04 & 1.89 (0.02) & 1.90 (0.01) \\ 
P05 & 1.81 (0.04) & 1.84 (0.02) \\
P06 & 1.85 (0.03) & 1.87 (0.02)\\ \hline
A02& 1.54 (0.09) & 1.55 (0.05) \\
A03 & 1.67 (0.05) & 1.72 (0.04) \\
A04 & 1.67 (0.05) & 1.71 (0.04)\\
A06 & 1.56 (0.07) & 1.64 (0.06) \\
A07 & 1.75 (0.06) & 1.80 (0.04) \\ \hline
\end{tabular}}
\end{center}
\caption{
Edge dimensions for selected Pollock (top) and Automatistes (bottom) images,
showing
higher complexity of luminance gradient patterns for greyscale distance
$\beta=1,30$.
}
\label{polautodims}
\end{table}

\pagebreak

\begin{table}[h]
\begin{center}
{\begin{tabular}{l c c l}\hline
Painting & $D_0 (\beta=1)$ & $\beta=30$ & Type \\ \hline
AV01 (XI) & 0.00 (0.00) &0.00 (0.00) & Homogeneous   \\
AV02 (XIII) & 0.00 (0.00) & 0.00 (0.00) & Homogeneous \\
AV03 (XV) & 0.00 (0.00) & 0.08 (0.05) & Homogeneous \\
AV04 (III) & 0.68 (0.03) & 0.83 (0.04) & Heterogeneous \\
AV05 (VII) & 0.95 (0.05) & 1.04 (0.05) & Heterogeneous \\
AV06 (VIII) & 1.28 (0.03) & 1.37 (0.04) & Heterogeneous \\ \hline
\end{tabular}}
\end{center}
\caption{
Effective edge dimensions for selected Avital images for greyscale distance
$\beta=1,30$.  Image sources
are identified by Plate number from \cite{avital}.
}
\label{avedges}
\end{table}

\pagebreak

\pagebreak

\pagebreak

\pagebreak

\pagebreak

\pagebreak

\pagebreak

\pagebreak

\pagebreak

\end{document}